# Development of Solar Flares and Features of the Fine Structure of Solar Radio Emission

G. P. Chernov[a,b,]*, V. V. Fomichev[b], Y. Yan[a], B. Tan[a], Ch. Tan[a], and Q. Fu[a]

[a] *Huairou Solar Observing Station, National Astronomical Observatory of China, Beijing, China* [b] *Institute of Terrestrial Magnetism, Ionosphere and Radio Wave Propagation, Russian Academy of Sciences, Moscow, Russia* *e-mail: gchernov@izmiran.ru*

**Abstract**—The reason for the occurrence of different elements of the fine structure of solar radio bursts in the decimeter and centimeter wavelength ranges has been determined based on all available data from terrestrial and satellite observations. In some phenomena, fast pulsations, a zebra structure, fiber bursts, and spikes have been observed almost simultaneously. Two phenomena have been selected to show that the pulsations of radio emission are caused by particles accelerated in the magnetic reconnection region and that the zebra structure is excited in a source, such as a magnetic trap for fast particles. The complex combination of unusual fiber bursts, zebra structure, and spikes in the phenomenon on December 1, 2004, is associated with a single source, a magnetic island formed after a coronal mass ejection.

## 1. INTRODUCTION

In the continuum emission of type IV solar bursts from meter to centimeter wavelength ranges, the following fine structure elements are usually observed: fast pulsations in a wide frequency range; fastest bursts of millisecond duration (spikes); narrow drifting bands in emission and absorption, among which there are fiber bursts with constant frequency drift and bands with various drift- a zebra structure (ZS) (Chernov, 2011). Due to the variety of unusual band shapes, the ZS is the most intriguing element of the fine structure. Therefore, it attracted special attention from researchers, and, after the first publication on ZS observations in the meter range (Elgaroy, 1959), more than a hundred works were published, not only on new ZS observations but also theoretical ones.

Over the past 40 years, more than a dozen emission mechanisms have been proposed to explain it. The most popular model is the emission mechanism under double plasma resonance (DPR) (Zheleznyakov and Zlotnik, 1975), in which the upper hybrid frequency $\omega_{UH}$ is compared with an integer number of electron cyclotron harmonics $\omega_{UH} = (\omega^2_{Pe} + \omega^2_{Be})^{1/2} = s\omega_{Be}$, where $\omega_{Pe}$ is the plasma frequency, $\omega_{Be}$ is the cyclotron frequency of electrons, and $s$ is the harmonic number provided that $\omega_{Be} \ll \omega_{Pe}$.

The interaction of plasma waves with whistlers can be considered an important alternative mechanism (Chernov, 1976): $l + w \rightarrow t$. Kuijpers (1975) proposed it to explain fiber bursts. However, in a source such as a magnetic trap, whistlers excited on the anomalous Doppler effect can be generated in the form of periodic wave packets. They propagate at a group velocity and determine the frequency drift of the ZS bands, which varies synchronously with the spatial drift of the radio source in the corona. Only the model with whistlers explains a number of fine effects of the dynamics of the ZS bands: the saw-tooth frequency drift, frequency splitting of bands, and hyperfine structure of bands in the form of millisecond spikes (Chernov, 2006).

A number of other ZS models (Treuman et al., 2011; Ledenev et al., 2006; Laptukhov and Chernov, 2006, 2012; Fomichev et al., 2009; LaBelle et al., 2003; etc.) have not yet been unequivocally recognized.

The relevance of ZS research has increased after the discovery of such bands in the radio emission of a pulsar in the Crab nebula under extreme physical parameters peculiar to pulsars (Zheleznyakov et al., 2012).

In many solar phenomena, all of the marked elements of the fine structure are present almost simultaneously on the dynamic spectrum, and their emission mechanisms can be different. The use of all available optical and X-ray data often helps in the flare analysis. Thus, in the phenomenon on April 11, 2013, the use of simultaneous images in several extreme ultraviolet lines from the Solar Dynamic Observatory/Atmospheric Imaging Assembly (SDO/AIA) helped to understand the repeated change in the sign of the circular polarization of the radio emission, when each new flare brightening occurred over regions with dif ferent magnetic polarities, and the ordinary wave mode remained (Chernov et al., 2016). Additional data in the case of the hard X-ray emission (RHESSI) made it possible to construct a probable radio source scheme within the standard model of a flare with magnetic reconnection, where the pulsations of radio emission in the 2.6–3.4 GHz band were associated with the upward acceleration of fast particles from the current sheet, and the ZS emission at lower frequencies was associated with the capture of particles accelerated downward, into a flare loop.

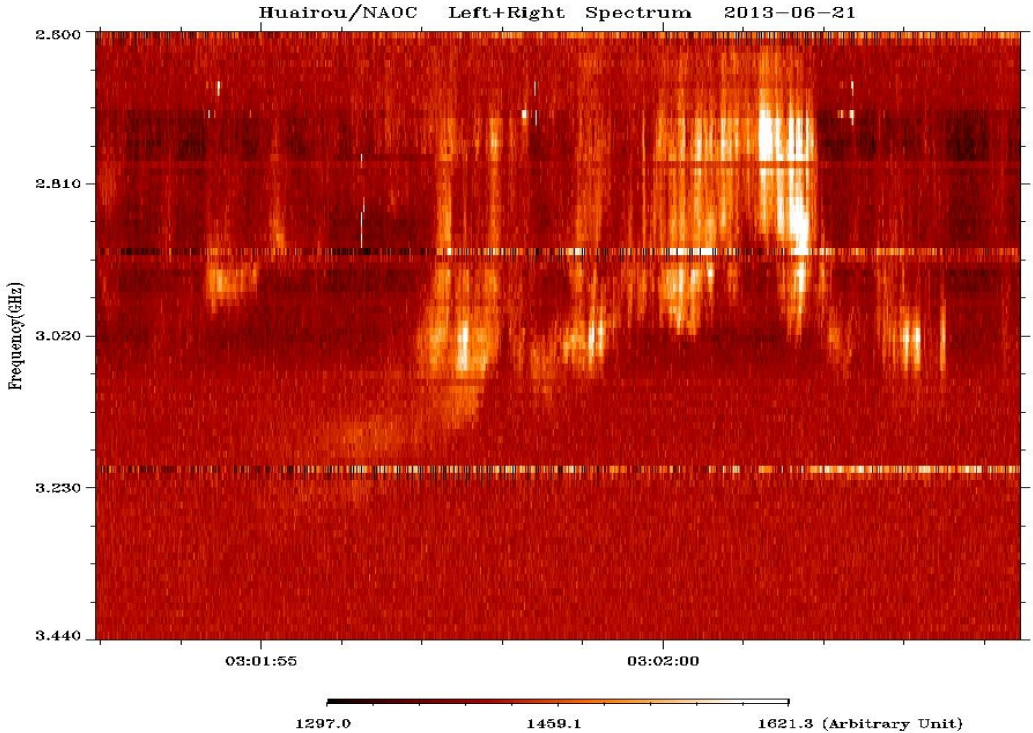

**Fig. 1.** Fast pulsations with an uneven emission cutoff boundary from the HF edge of the spectrum if there is no ZS.

However, in the following phenomenon on June 21, 2013, a group of fast pulsations also developed around 3 GHz, but there was no ZS. In the phenomenon on December 1, 2004, on the contrary, from the very beginning of the phenomenon, there were different fiber bursts, spikes, pulsations, and ZS almost simultaneously on the spectrum in the decimeter range 1.1–1.34GHz.

In this paper, it is attempted to understand this hierarchy of fine structure and to relate it to the dynamics of the flare process. The statistical analysis of the ZS is complicated by a wide variety of phenomena (Tan et al., 2014). In the absence of high-resolution positional observations of radio sources, it is first important to understand the causes of the sequential appearance of individual elements of the fine struture, and it is even more important to understand their simultaneous appearance, since no such analysis of phenomena has been conducted so far.

## 2. NEW OBSERVATIONS

The paper uses data from solar broadband radio spectrographs (SBRS) of the National Astronomical Observatory of China (NAOC) installed at the Huairou station near Beijing: for the phenomenon on June 21, 2013, the spectrograph in the centimeter range 2.6–3.8 GHz with a frequency resolution of 10 MHz and the time resolution of 8 ms, and for the phenomenon on December 1, 2004, the spectrograph in the decimeter range 1.1–1.34 GHz with a high resolution of 4 MHz and 1.25 ms (Fu et al., 2004). For an analysis of phenomena in general, all available satellite data were used: SOHO/LASCO C2, SOHO/MDI/EIT, RHESSI, and SDO/AIA.

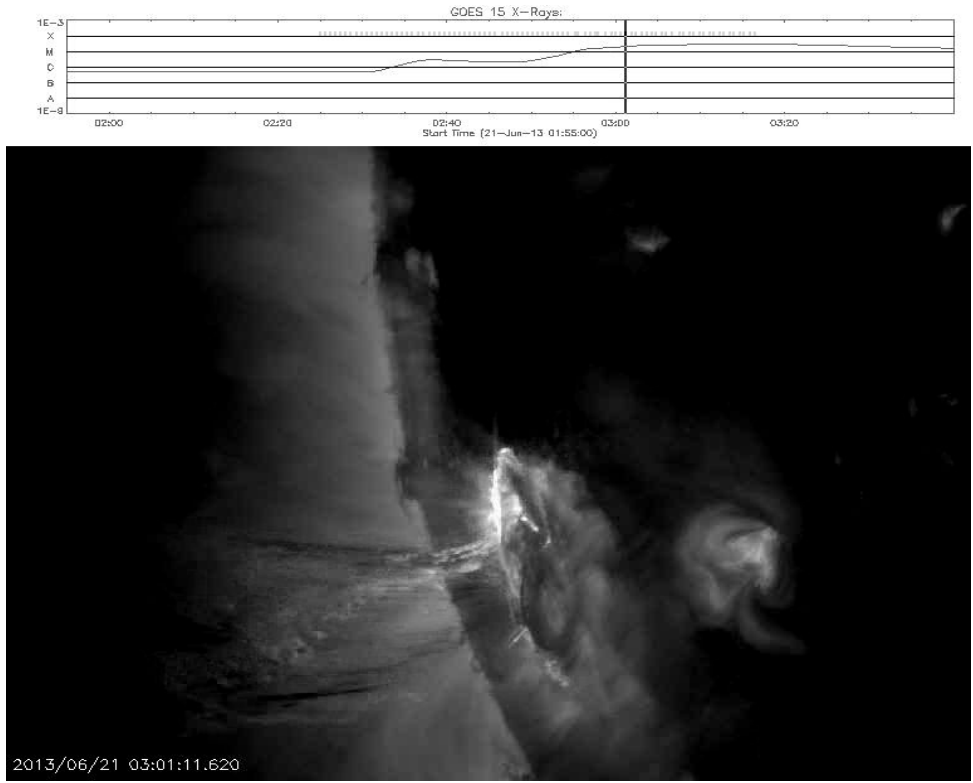

**Fig. 2.** SDO 211 Å. Ejection along the open field lines goes high into the corona. There is no magnetic trap for particles. Soft X-ray emission profile from the GOES 15 (top panel). Start time is January 21, 2013, 01:55:00.

### 2.1. Phenomenon on June 21, 2013

This phenomenon is remarkable in that pulsations and ZS appeared independently at different instants. A 2.9 f lare occurred in the eastern region of NOAA 11777 (12 S, 75 E). The first fast pulsations in the 2.6–3.2 GHz range appeared right after the coronal mass ejection (CME) at 03:01:50 UT but without ZS bands from the high-frequency (HF) edge, in contrast to the phenomenon on April 11, 2013, mentioned in the INTRODUCTION (Fig. 1).

The CME starts at 02:45 UT. In SDO/AIA films, it starts with an explosion of a dark filament. The emission goes high into the corona along the open field lines (Fig. 2), and, by the time of the pulsations at 03:01:50 UT, the last small emissions in the tail of the CME (Fig. 3) are still visible. At the same time, the flare brightening begins. It is probably related to the magnetic reconnection and, consequently, to the acceleration of particles, as a result of which fast pulsations occur (Fig. 1). As can be seen in Fig 3, there are no closed loops (magnetic trap) at this time, and therefore there are no conditions for the ZS excitation (within DPR or whistler models).

However, the ZS appeared about 1000 UT (Fig. 4). Exactly at this time, the formation of the flare loop can be observed (Fig. 5), probably after the repeated brightening of the flare visible on the film. After about three minutes, the flare loop fades.

An even more intense ZS appeared only at 03:25:25 UT (Fig. 6). At this time, the new flare brightening ends with the appearance of a new flare loop in the same place. As can be seen in Figs. 5 and 7, the height of the flare loops corresponds to a centimeter range of the plasma frequency.

In both cases, the ZS radio emission had a very weak polarization, which is characteristic of limb phenomena. In addition, it can be seen that the maximum energy release occurred at the tops of the loops. Thus, it can be clearly seen that the ZS appeared only at the time of the formation of flare loops, magnetic traps for fast particles.

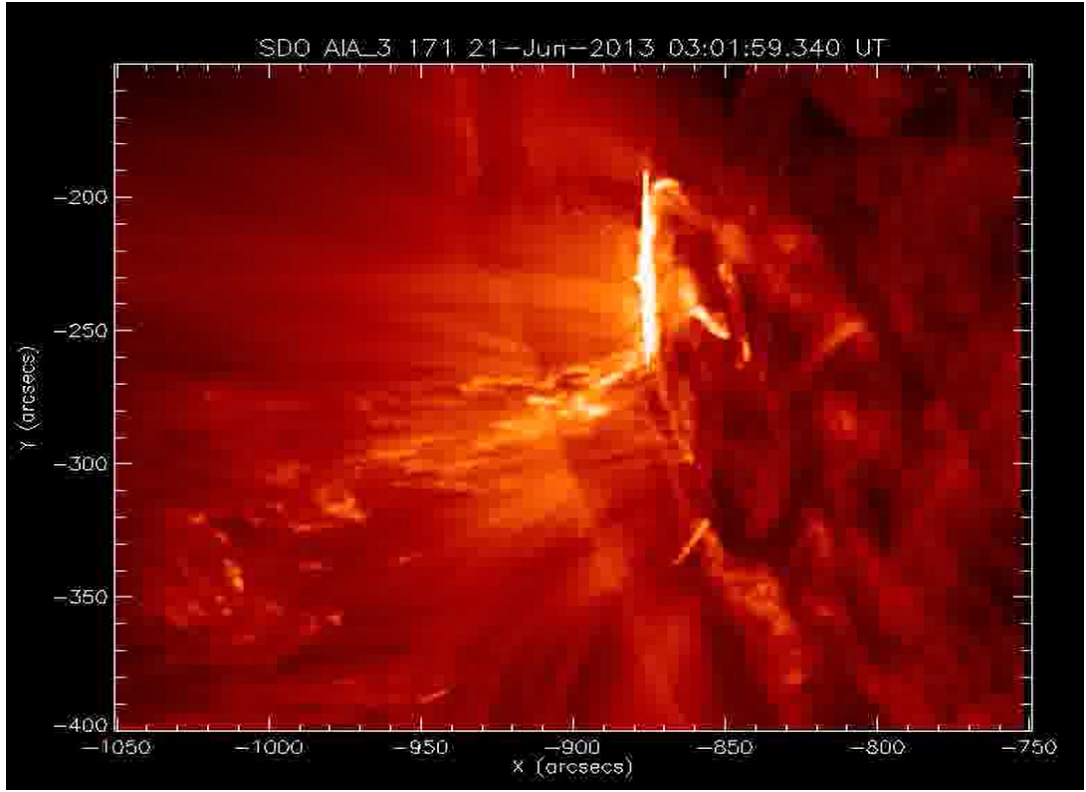

**Fig. 3.** SDO 171 Å at the time of pulsations. Last CME traces along the open field lines. Flare brightening indicates particle acceleration, but there is no magnetic trap, there are no conditions for the ZS generation, and, thus, only pulsations are observed.

### 2.2. Phenomenon on December 1, 2004. Observations

The phenomenon on December 1, 2004, has been discussed for more than 10 years, but each work has been devoted to a single selected effect. While Huang et al. (2007) considered the phenomenon as a whole, including an analysis of the fine structure of the radio emission (however, only on the pulse burst phase) (Ning et al., 2009; Gao et al., 2014), two unusual fiber burst groups at the onset of the phenomenon were analyzed (Fig. 8). In the first paper (Liu et al., 2006), in which only these unusual fiber burst groups were mentioned, they were classified as fantastic patterns of the fine structure of bursts that resemble a hand with fingers outstretched (the term used in (Allaart et al., 1990)). The ZS and the so-called lace bursts were observed throughout the phenomenon for more than 30 min (Huang and Tan, 2012).

The onset of the phenomenon is described in the literature (Huang et al., 2007). A small M1.1 flare occurred in the active region of NOAA 10708 near the center of the disk (06 N, 20 E). It started at 07:06, reached the maximum at 07:15, and proceeded to 08:21 UT according to X-ray emission data (in soft emission, GOES, and in hard emission, RHESSI). Even before the maximum, five separate peaks were observed. In the mentioned papers, the evolution of the fine structure and the mechanisms of radio emission were not analyzed.

After two groups of unusual fiber bursts (Fig. 8), fiber bursts immersed in the developed ZS were observed (Fig. 9). The ZS lasted about a minute and ended with the limitation by HF fiber bursts (Fig. 10). A similar ZS limitation with fiber bursts from the HF edge was observed later in the microwave range 2.6–3.8 GHz in the phenomenon on August 1, 2010 (Chernov et al., 2014). The circular polarization sign changed seven times within 2.5 min (Fig. 11). Such polarization changes are consistent with the dynamics of sources in hard X-ray emission (Fig. 12): each new f lare brightening coincided with a new source (*A*, *B*, *C*, *D*) in different energy ranges. The first such brightening above the tail sunspot was observed at 07:05:55 UT and in the extreme ultraviolet line 284 Å at the SOHO/EIT (Fig. 13).

Ning et al. (2009) associate fine-structure emission with fast particles accelerating downward from the magnetic reconnection region. At the beginning, two groups of fiber bursts are associated with slow downward ejections, which explain the slow positive drift of the fiber bursts. One can only guess that the narrow frequency band of the fiber bursts is associated with the size of these emissions. The ZS is considered in DPR and whistler models, in which the source is assumed to be in the same place of the flare loops where fiber burst sources were located several seconds earlier.

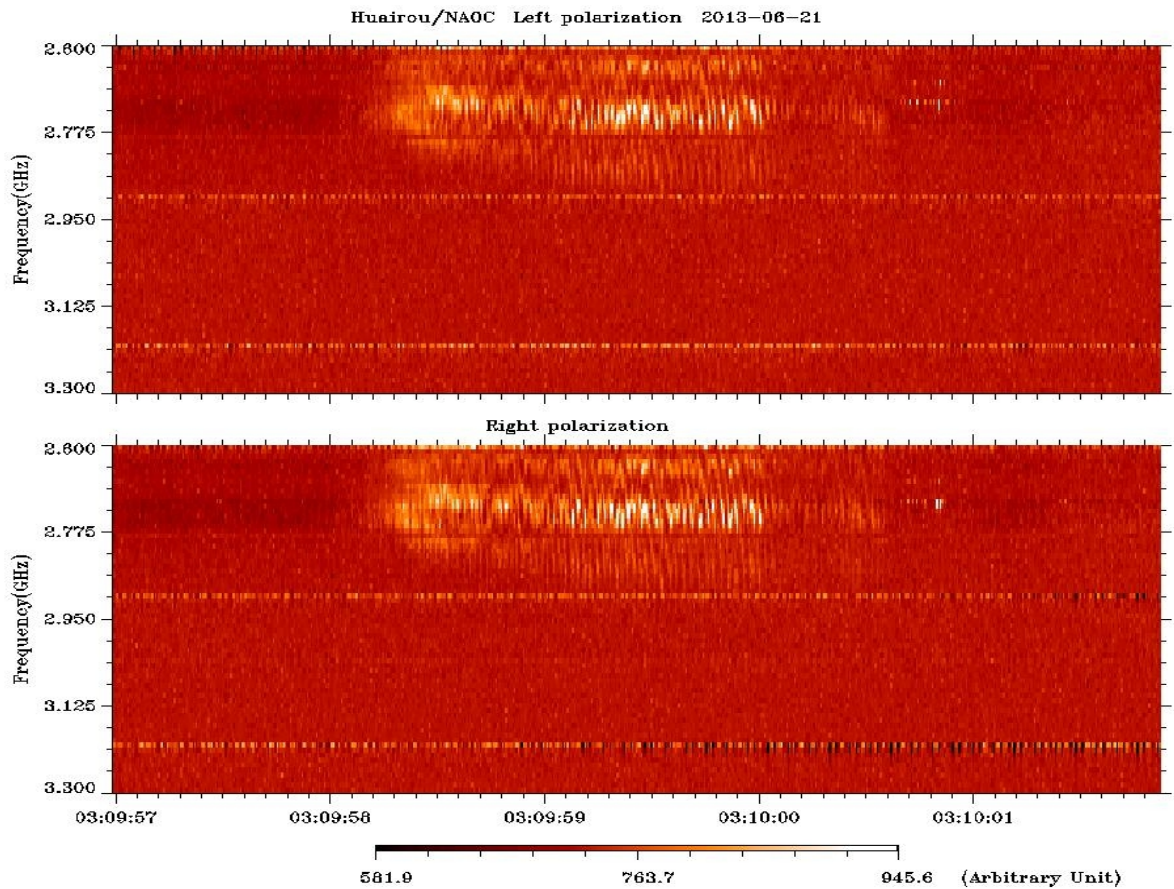

**Fig. 4.** First ZS appearance in the phenomenon on June 21, 2013. SDO AIA_3 171 June 21, 2013 03:09:59.340 UT

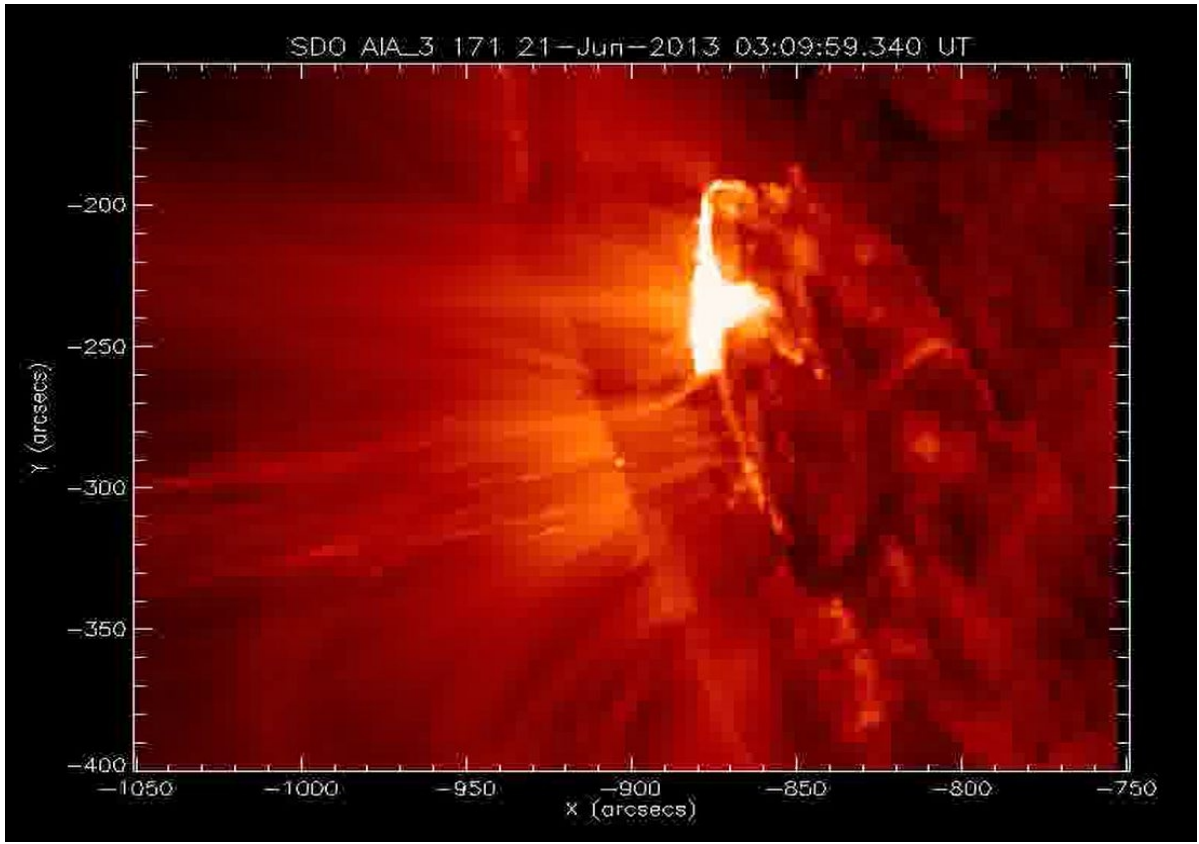

**Fig. 5.** SDO 171 Å. Flare loop was formed at the time of the appearance of the first ZS.

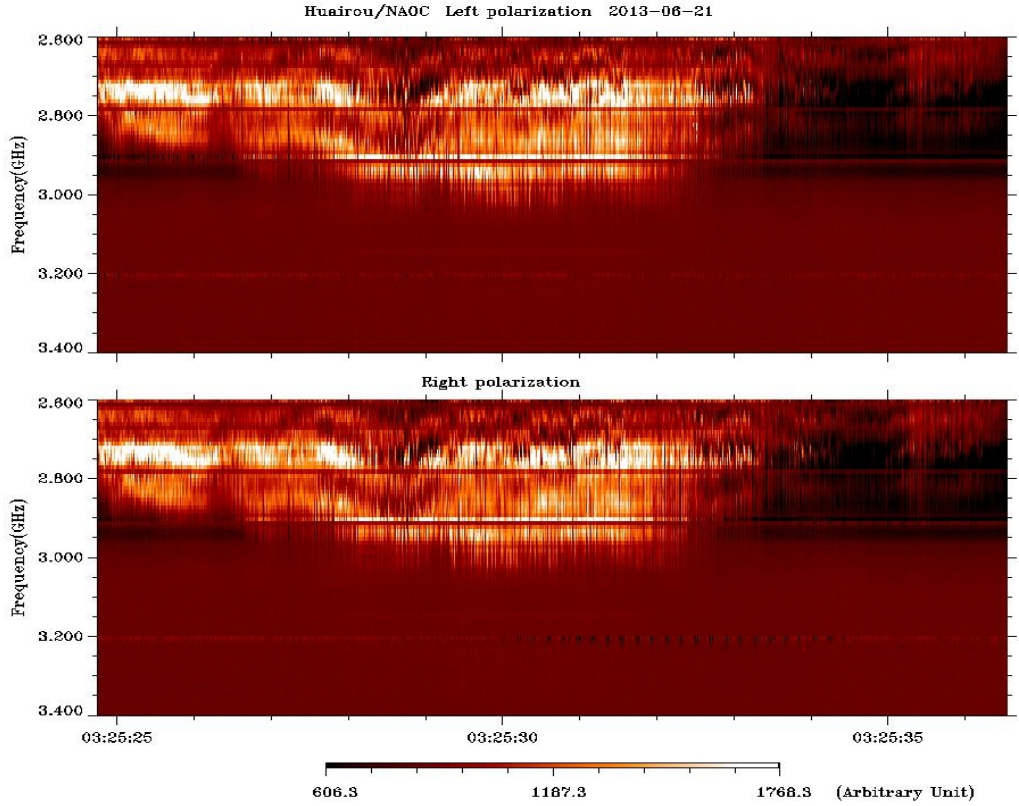

**Fig. 6.** Second ZS appearance at the end of the phenomenon on June 21, 2013.

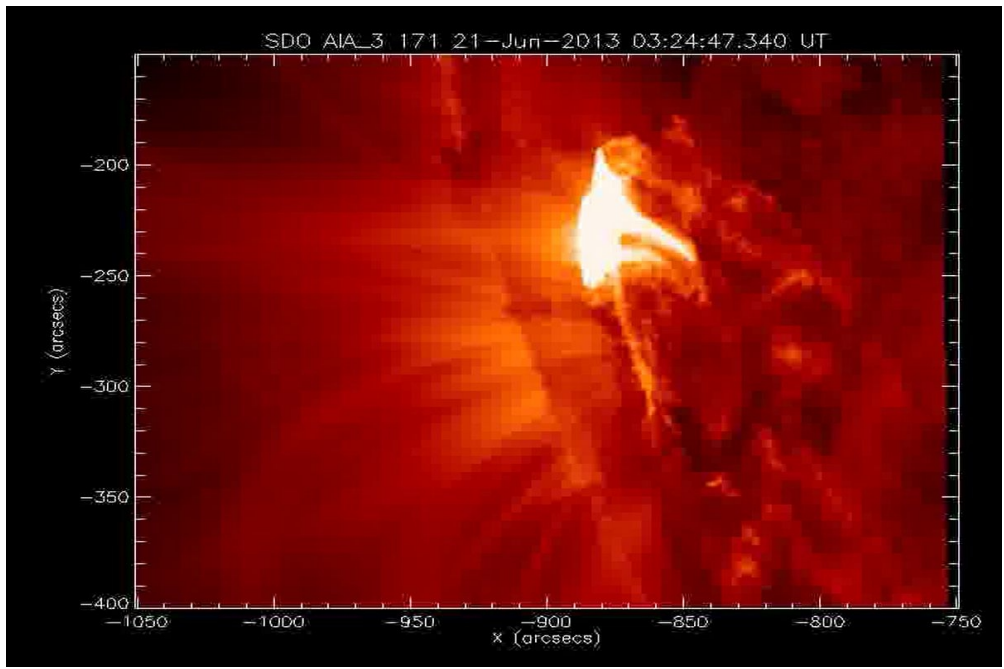

**Fig. 7.** SDO 171 Å. Flare loop in the form of a magnetic trap appeared just at the time of the appearance of the ZS at the end of the phenomenon.

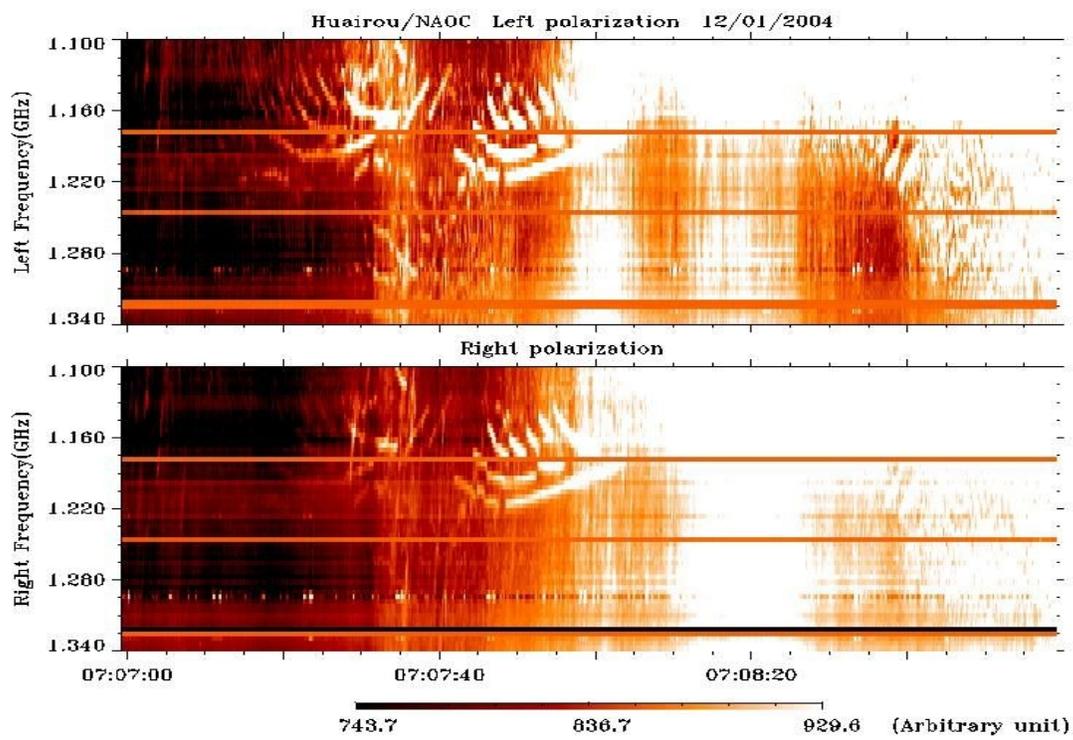

**Fig. 8.** Onset of the phenomenon on December 1, 2004, with a duration of 1 m 40 s. Moderate left-handed polarization of the fiber bursts and weak right-handed polarization of the continuum.

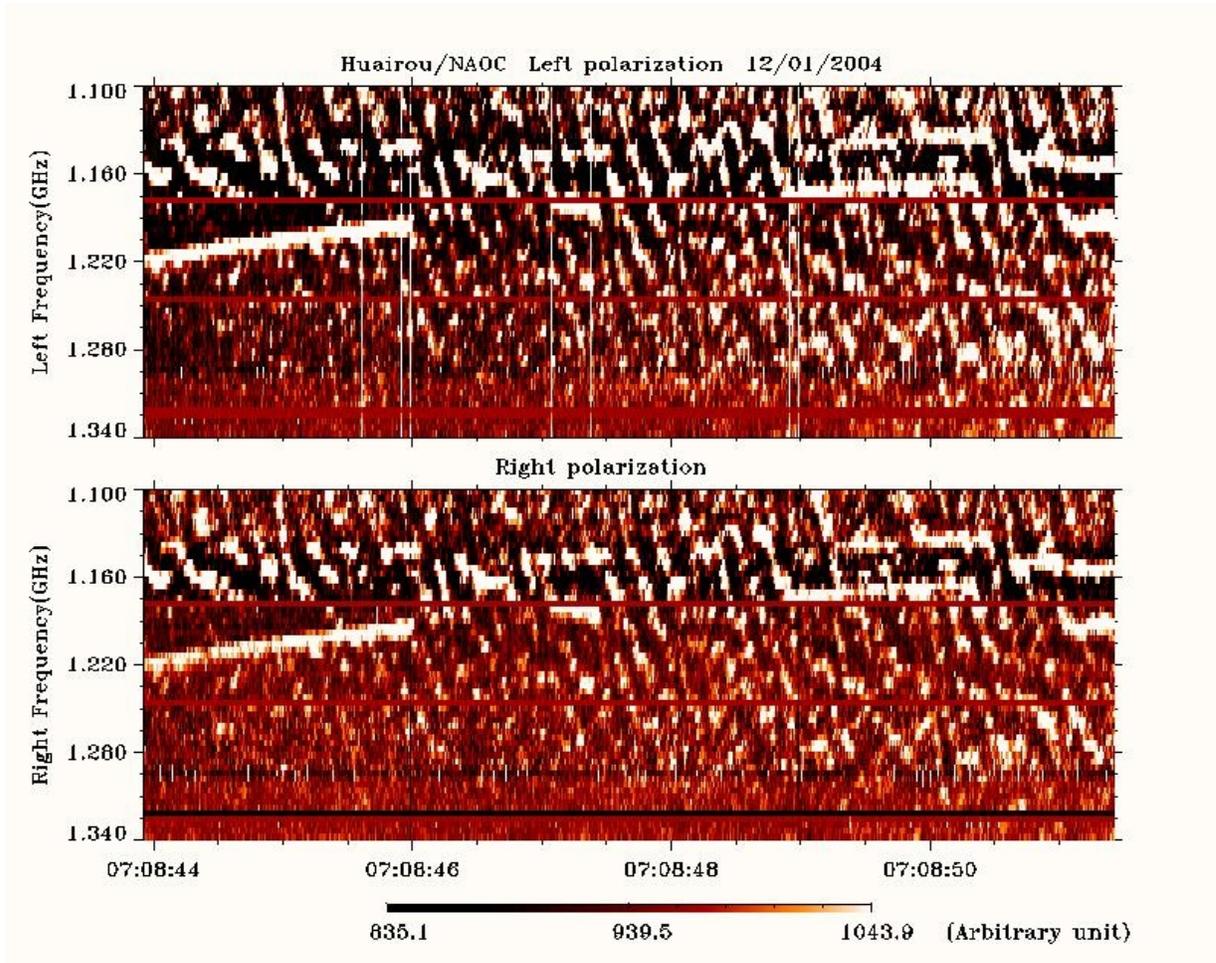

**Fig. 9.** Continuation of the phenomenon on December 1, 2004 (duration of 8 s) with the developed ZS with fiber bursts immersed in it. The emission has a moderate (weak) right polarization (see the temporal profiles below).

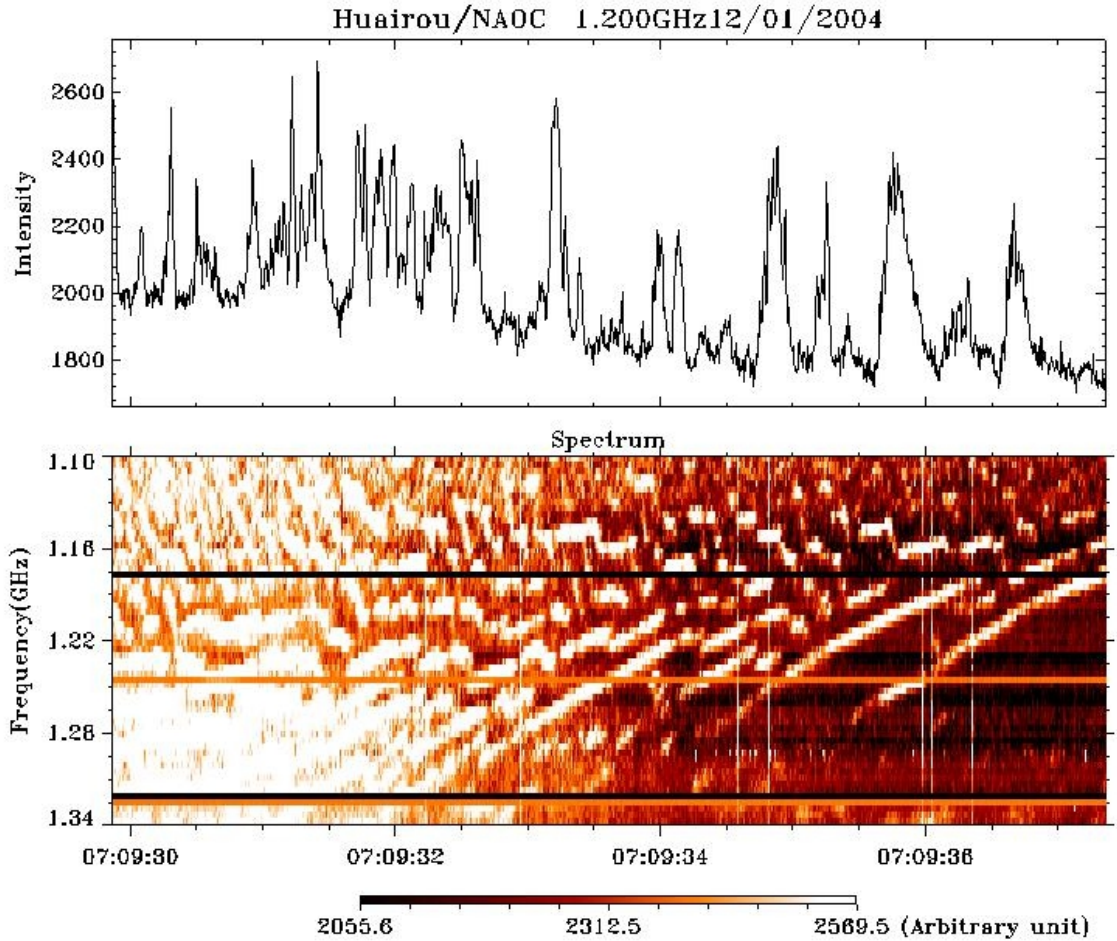

**Fig. 10.** Zebra structure is limited from high frequencies by a series of fiber bursts in the phenomenon on December 1, 2004, in the range 1–1.34 GHz. Right polarization was changed to the left again.

Ning et al. (2009) associate two unusual groups of fiber bursts at the onset of the phenomenon with the chromospheric evaporation as a result of its bombardment by fast particles accelerated downward from the magnetic reconnection region. However, it can only be considered as an assumption, since there is no temporal coincidence of evaporation instants (according to the hard X-ray data) with the occurrence of the fiber bursts. In addition, there is no explanation for the narrow frequency band of the fiber bursts, their frequency drift, and periodicity.

Note that the most real explanation for unusual fiber bursts at the onset of the phenomenon was proposed in (Gao et al., 2014). The emission of fiber bursts was associated with the ejections of magnetic islands (clouds) from the region of magnetic reconnection and the rise of flare loops. However, the mechanism of the generation of fiber bursts was not analyzed. The possible role of termination shock was indicated, behind the front of which particle acceleration must occur.

### 2.3. Interpretation of the Phenomenon on December 1, 2004

The beginning of the CME on December 1, 2004, is projected (approximated) at 07:00. This is the begin-ning of the flare in the soft X-ray at GOES. Therefore, the CME was caused by the initial explosion of the filament. However, there was no type II burst. The HIRAISO spectrograph only gives a type IV burst. At the beginning, there was only one HXR 5–6 keV source over the head sunspot, and it became double (at the bases of the loops) by 07:05 UT. This is the effect of the fast particles that reached dense plasma at the bases of the loops, which was used as evidence of chromospheric evaporations (Ning et al., 2009). After the departure of the CME, these particles could be accelerated during a prolonged magnetic reconnection with the formation of the magnetic island. The magnetic island probably already existed by the time 07:07:30 UT, and its size determined the frequency range of two groups of unusual fiber bursts. Therefore, we propose a new flare scheme with a magnetic island (Fig. 14). The particles are accelerated in the current sheet at the bottom of the magnetic island, not only downward but also upward. Since the onset of particle acceleration, when there was no continuum, the upward accelerated particles should excite plasma waves and whistlers. Whistlers excited on the normal Doppler effect must propagate downward. Thus, unusual groups of fiber bursts with positive frequency drift are fiber bursts excited by wave packets of whistlers trapped in a magnetic cloud. They cannot spread below the reconnection point. The frequency drift is slowed down by this capture.

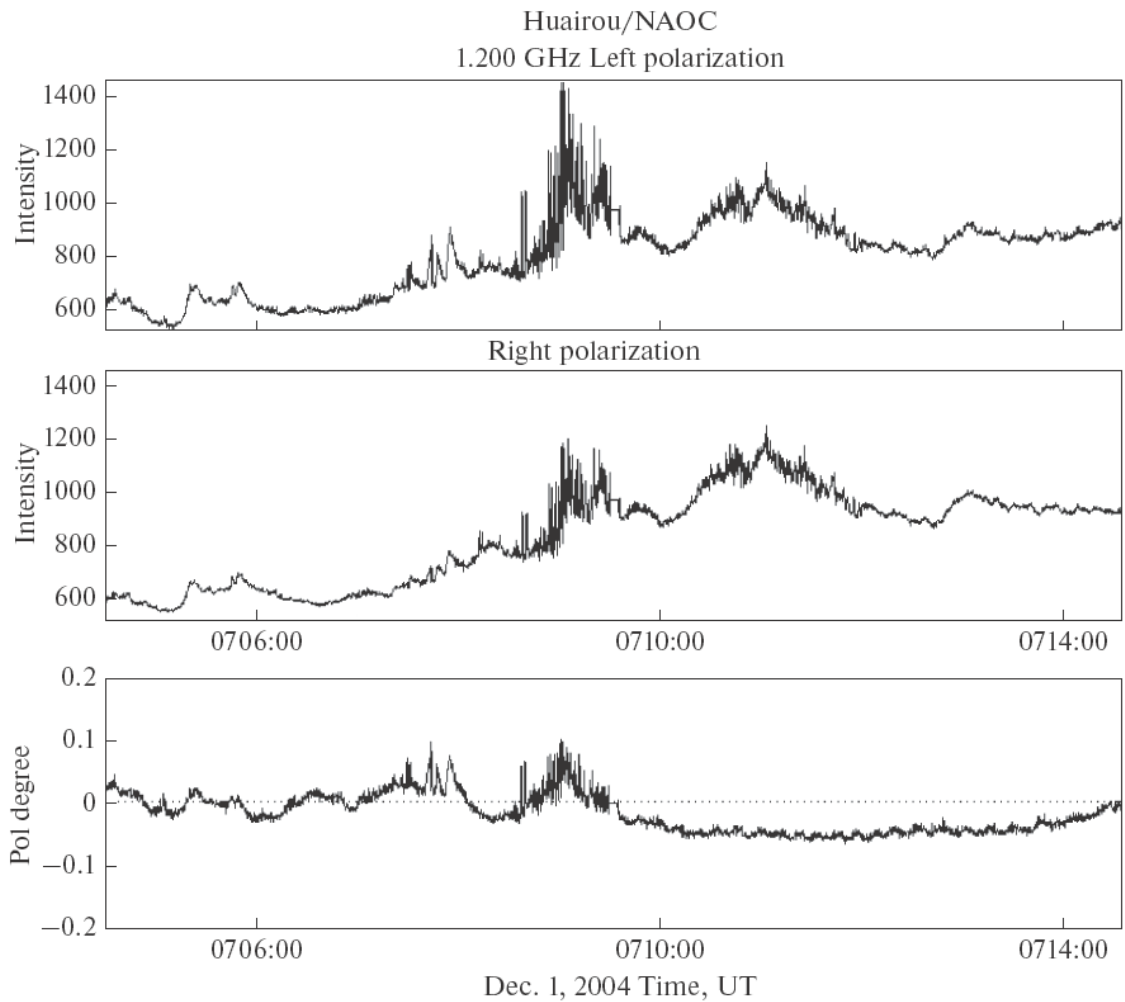

**Fig. 11.** Time profiles of channels in left and right polarization and degrees of polarization. Polarization changed the sign seven times.

Particles accelerating downward excite whistlers that propagate upward. They cause fiber bursts with negative frequency drift. Therefore, on the spectrum they look like a high-frequency boundary of groups of fiber bursts with positive drift. Simultaneously, the cloud could first descend and meet a closed flare loop. Obvious plasma ejections down from the current sheet should cause a shock wave that encounters a barrier in the form of a flare loop. Therefore, the shock wave should inevitably become a termination shock (TS), in which the particles are additionally actively accelerated and which was mentioned in the literature (Gao et al., 2014). In this paper, the authors are doubtful only because of the fact that they have no other proof of the existence of the termination shock. However, a recent paper (Chen et al., 2015) provides such evidence for another phenomenon.

The first effect of particles accelerating in the TS front is fast spikes like primary energy release (Chen et al., 2015). Some of the particles are captured in the magnetic cloud, where the velocity distribution with a loss cone (or a ring distribution) is formed; the ZS emission is associated with this. In the first seconds, the particles (before capture) should trigger episodic fiber bursts immersed in the developed ZS.

All of the above effects from particles accelerated in the TS front can be seen in Fig. 15. The fiber burst at the beginning of the upper spectrum actually looks like a continuation of fiber bursts with negative frequency drift limiting two groups of unusual fiber bursts from high-frequency bands at the beginning of the phenomenon. However, it is already immersing in a family of spikes and after six seconds it is divided into several fiber bursts, which transform after 5 s to several ZS bands with a wave-like drift. The particles and whistlers gradually fill the entire magnetic cloud, and on the third tab of the spectrum we observe the expansion of the ZS frequency band almost to the entire range, which is only partially surrounded by spikes. On the lower tab, the ZS is mixed with spikes (or spikes are organized into ZS bands, which in part look like fiber bursts with negative drift). All of these processes occur in a rather narrow frequency band of 1.1–1.34 GHz, in a magnetic island source.

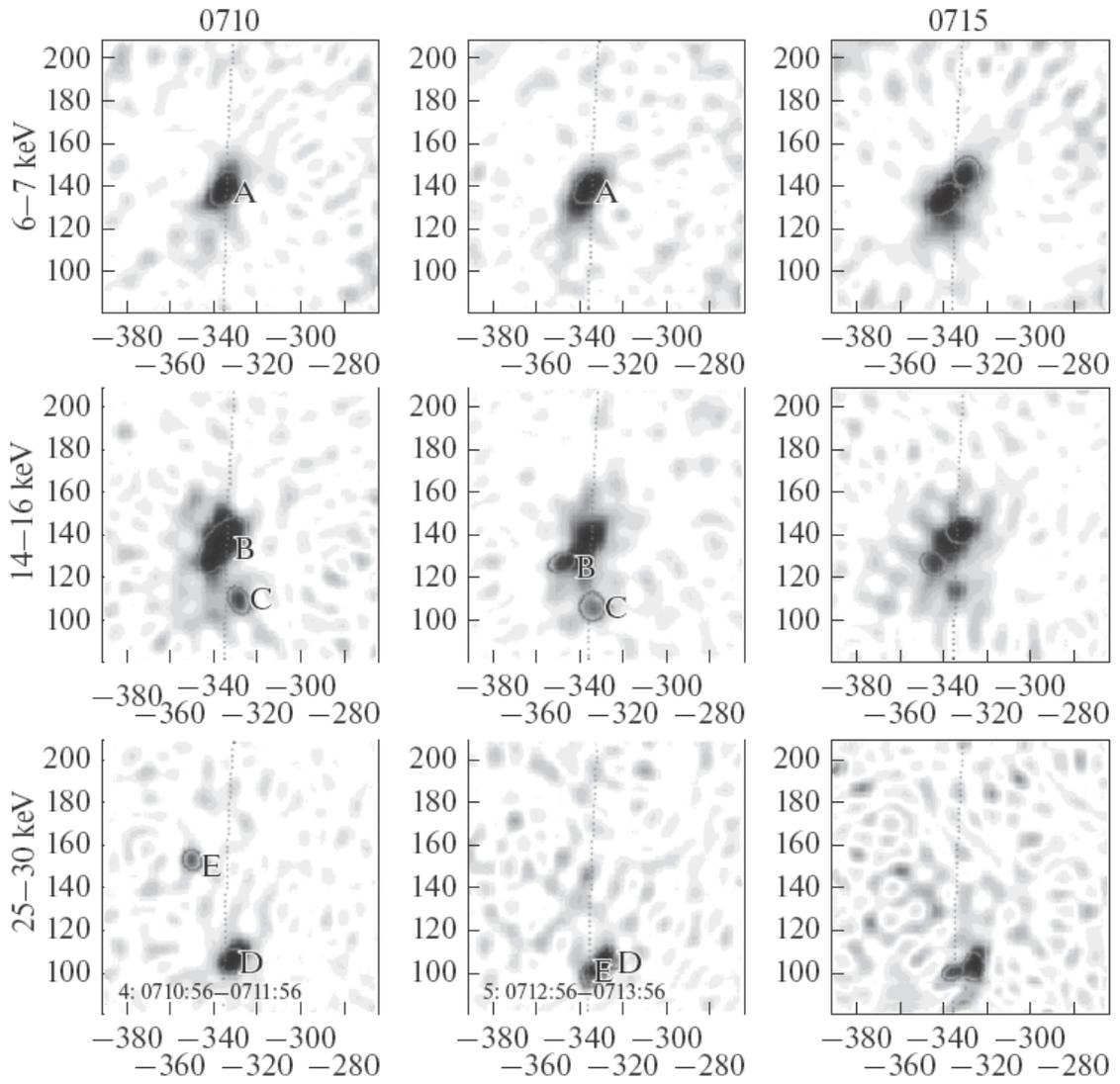

Fig. 12. Dynamics of hard X-ray sources (RHESSI).

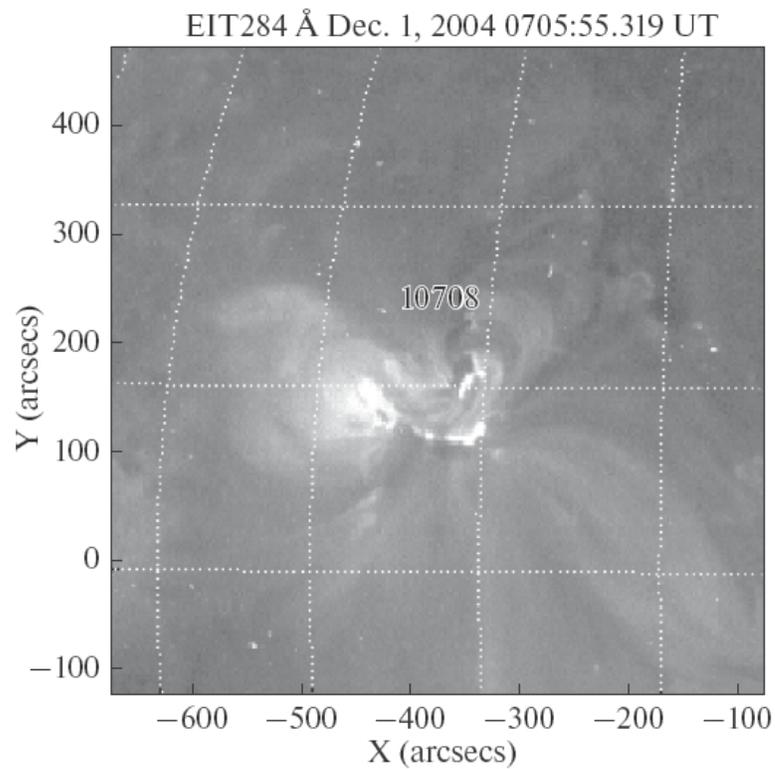

**Fig. 13.** Flare brightening in line 284 E (SOHO/EIT) over the tail sunspot at 0705:55 UT of the northern magnetic polarity. There is apparently just not one, which is the reason for the multiple polarization reversal.

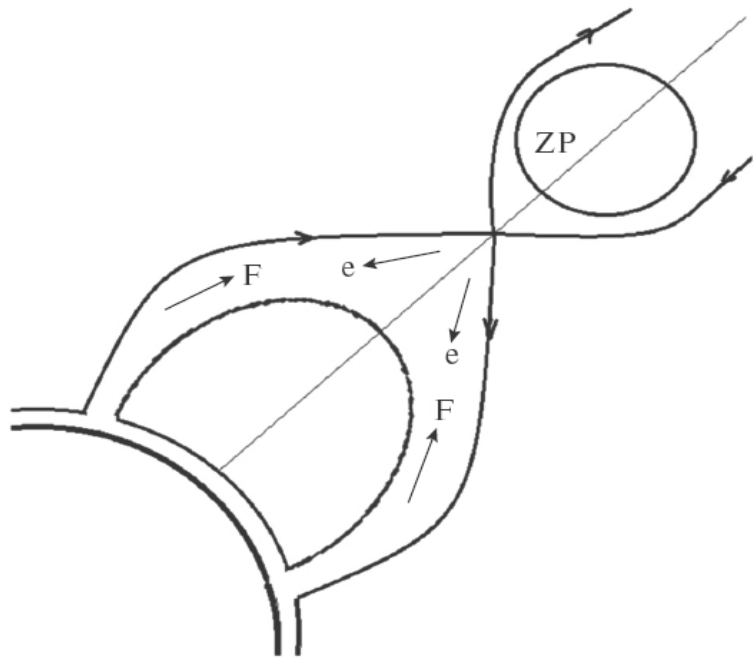

**Fig. 14.** Scheme of the flare for the phenomenon on December 1, 200 4. This is a long-lasting phenomenon. Magnetic islands form above the flare loop. The scheme refers to the instant after the interaction of the island with the flare loop with the formation of the $X$ point of magnetic reconnection and acceleration of the particles in both directions from the current sheet.

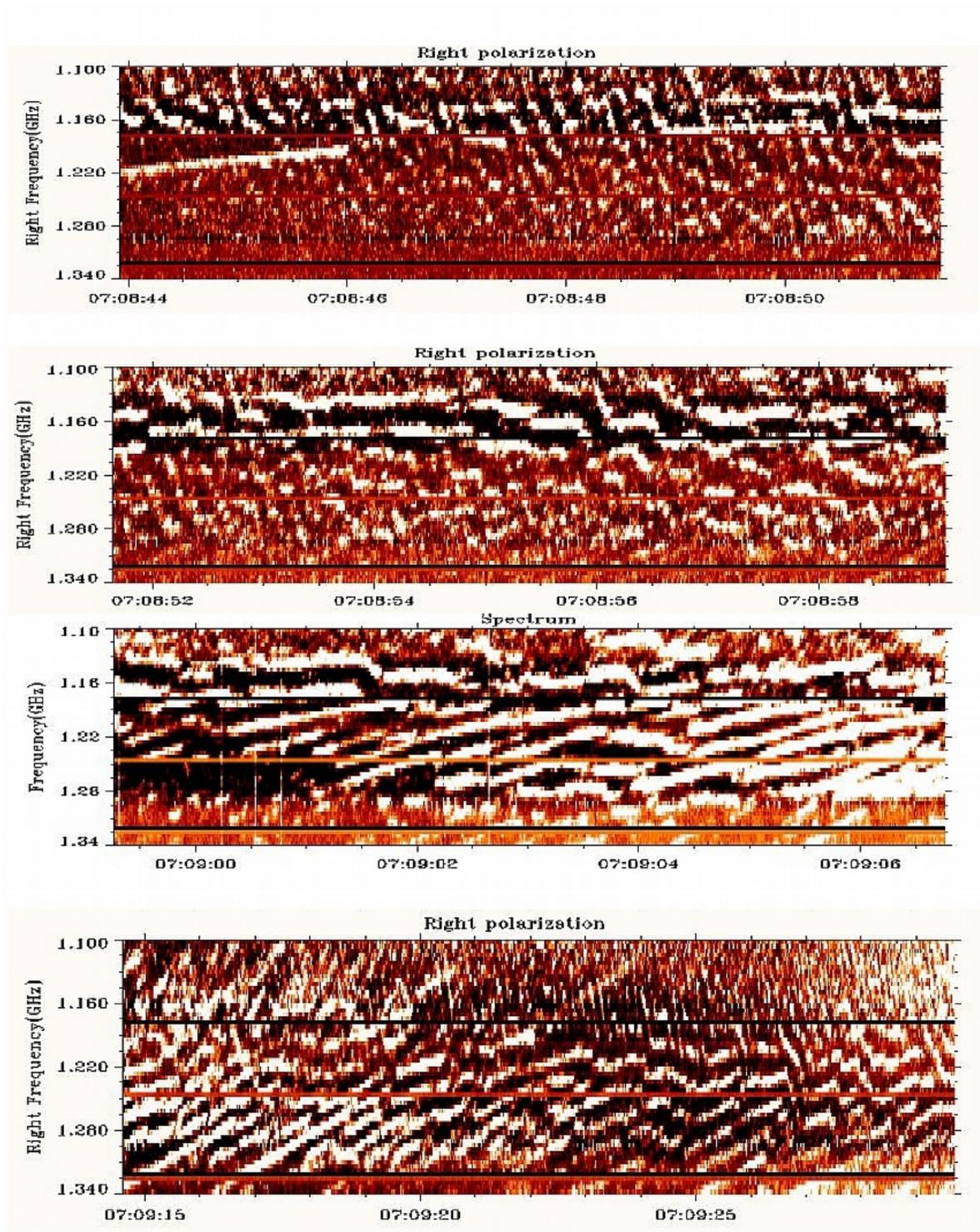

**Fig. 15.** Gradual transformation of fiber bursts into zebra and its evolution within 46 s.

The polarization of the continuum changed sign seven times during the first 10 min of the burst (Fig. 11). Even a cursory glance at the images of hard X-ray sources (Fig. 16) is enough to understand the effect of the alternating predominance of the northern (with the left sign of polarization) and southern (with the right sign of polarization) sources. The 25–30 keV HXR source appeared on the side, in the southern part of AR 10708. After ~07:11 UT, there was only one southern source for several minutes (C in Fig. 12), and the polarization at that time preserved the right sign. According to MDI's AR magnetograms (Fig. 6, (Ning et al., 2009)), radio emission always corresponded to the ordinary mode.

Later, up to 07:17 UT, episodic ruptured fiber bursts appeared (Fig. 17) against the background of a continuum consisting of spikes with a duration at the instrument resolution limit of 1.25 ms. That is, the acceleration in the TS front continued. It is difficult to say whether the turbulence of the plasma inside the magnetic island increased or, conversely, decreased. However, fiber bursts and clouds of spikes indicate the presence of whistlers and ion-acoustic waves. It is possible that there was no rapid rupture of the TS (as in the phenomenon of March 3, 2012 (Chen et al., 2015)). Most likely, the TS front along with the top of the flare loop experienced slow shifts (oscillations) up and down, and the particles and waves continued to be captured in the magnetic island. The gradual evolution of the fine structure (the initial unusual fiber bursts in the ZS, spikes, new fiber bursts, and lace bursts) is associated with the evolution of the distribution function of fast particles inside the magnetic island.

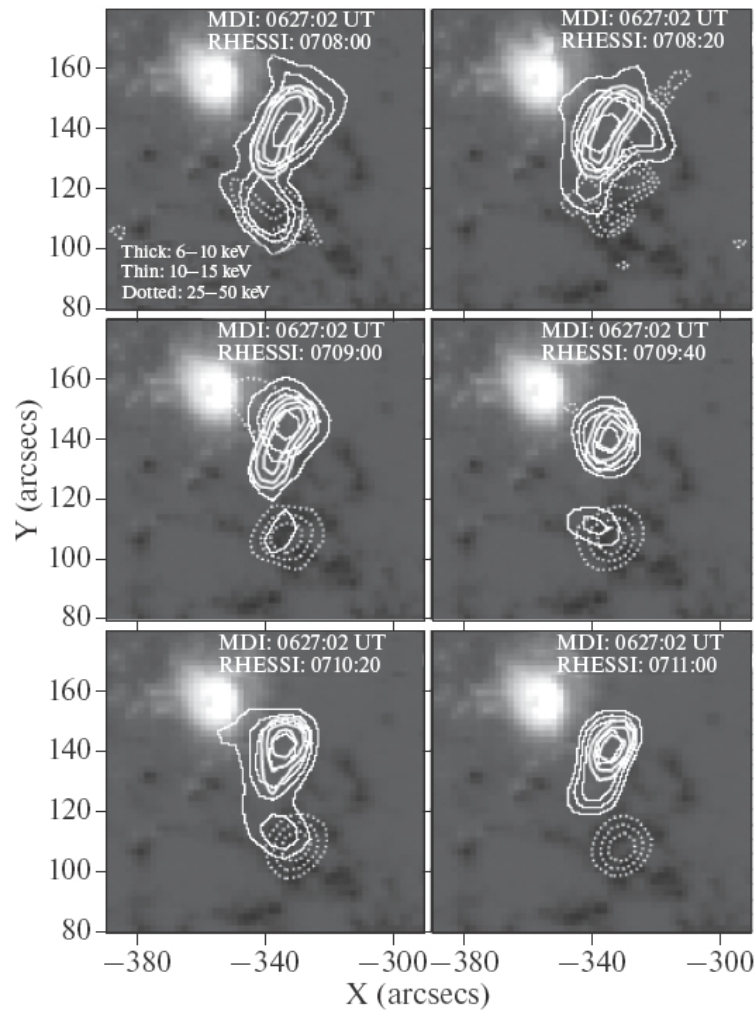

**Fig. 16.** Hard X-ray contours (RHESSI) against the background of a magnetic map (MDI) for six instants of the phenomenon on December 1, 2004. The levels of X-ray contours correspond to 40, 60, and 80% for 25–50 keV (dotted line), 50, 60, 70, and 90% for 10–15 keV (thin lines), and 70, 80, and 90% for 6–10 keV (thick lines). (Fig. 6 in (Ning et al., 2010)).

The fine structure at the damped phase of the burst was analyzed (Huang and Tan, 2012), and special attention was paid to the so-called lace bursts, which appeared at 07:25 UT (Fig. 18) and continued to 07:33:30 UT.

Lace bursts were interpreted (Karlický et al., 2001) as the radiation from turbulent plasma under conditions of double plasma resonance, when the paramters of the magnetic field and density change abruptly and often even violate the DPR condition. It is difficult to expect sharp fluctuations of the magnetic field and density at the damped phase of the f lare. Huang and Tan (2012) associate lace bursts with the emission of Bernshtein modes, although there is little evidence for this, both because of the chaotic form of the bands and because of the strong polarization of the bursts. However, estimates of the strength of the magnetic field by Bernstein modes coincided with the values obtained for fiber bursts and the ZS, ~ 70 Gs.

In this phenomenon, it was more obvious to associate the lace bursts with the wave packets of whistlers trapped in the magnetic island, since throughout the phenomenon the fiber bursts transforming into ZS bands and back are clearly associated with the whistlers.

## 3. CONCLUSIONS

In the absence of high-resolution positional observations of radio sources, we tried to understand the causes of the sequential appearance of individual elements of the fine structure and to explain their simultaneous appearance in the decimeter and microwave wavelength ranges, since no such analysis of the phenomena has been conducted so far. All available terrestrial and satellite observations were used. Two phenomena were selected to show that the pulsations of radio emission are caused by particles accelerated in the magnetic reconnection region, and the zebra structure is excited in the source such as a magnetic trap for fast particles. The complex combination of unusual fiber bursts, ZS, and spikes in the phenomenon on December 1, 2004, is associated with the development of instabilities in one source, a magnetic island formed after a coronal mass ejection.

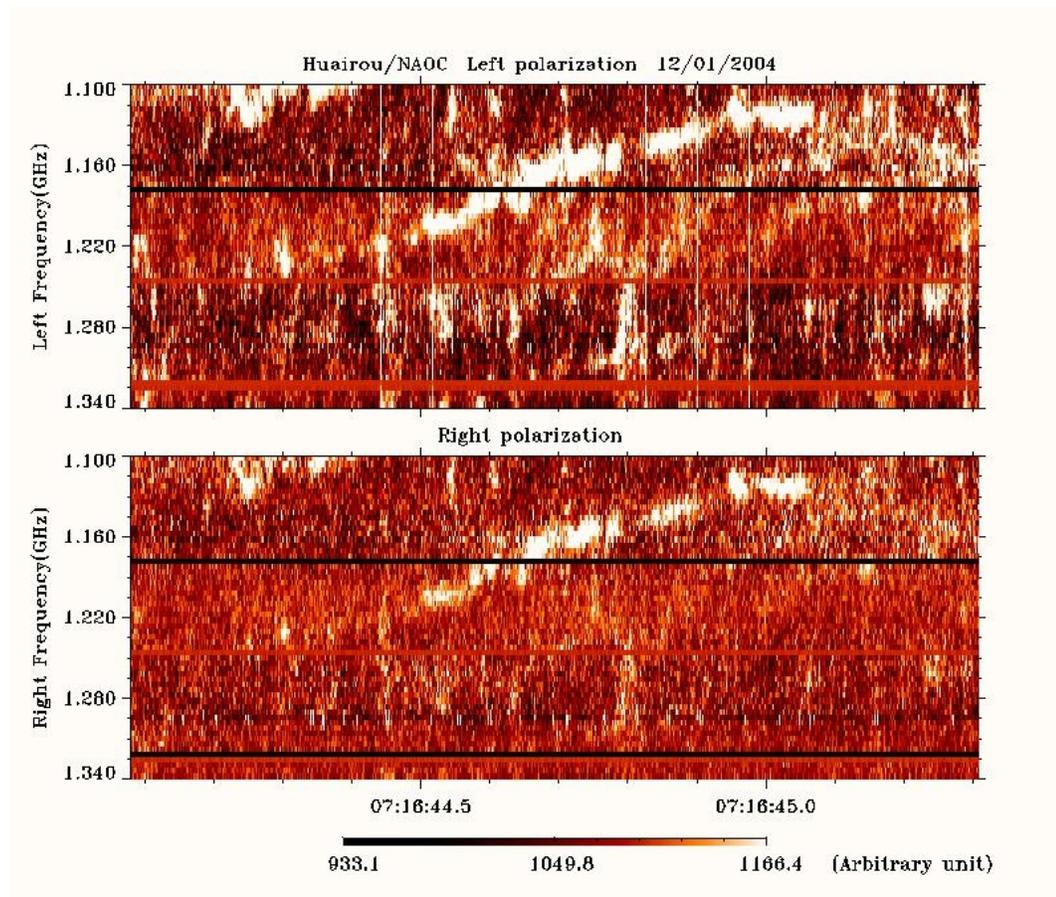

**Fig. 17.** Ruptured fiber bursts in the decreasing phase of the flare.

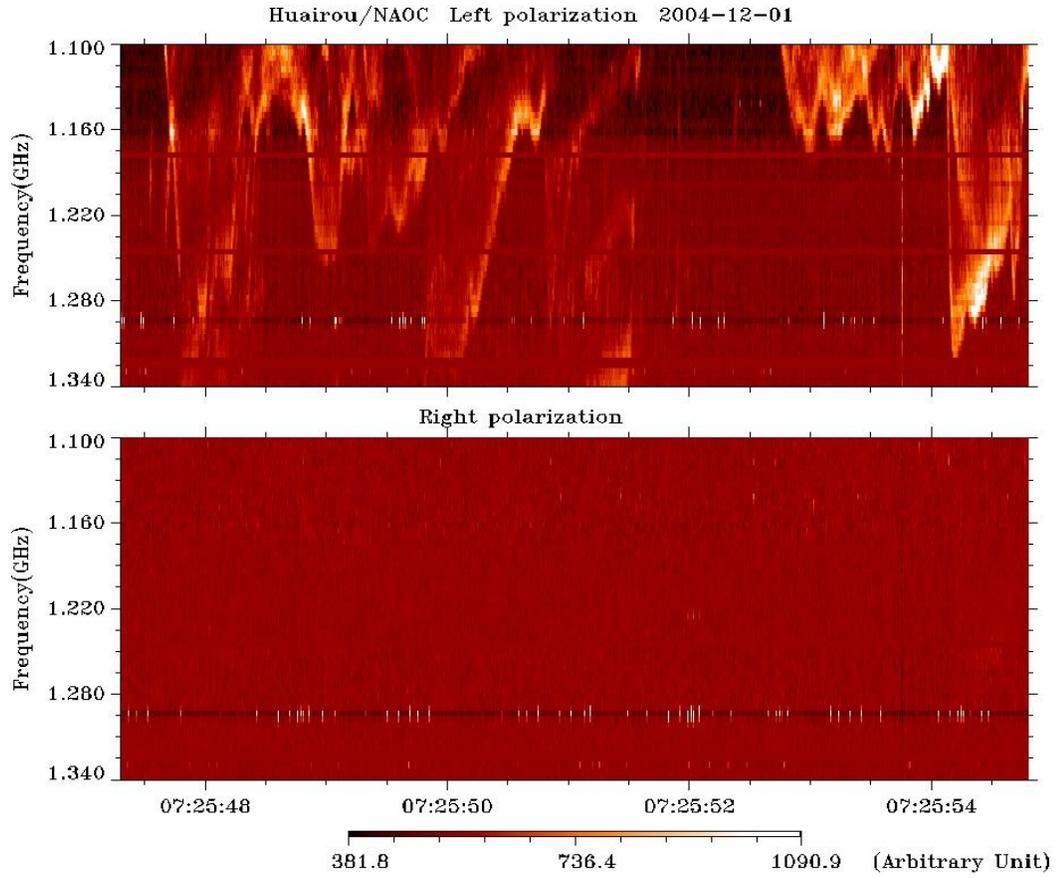

Fig. 18. Lace bursts in the decreasing phase of the flare.

ACKNOWLEDGMENTS

We are grateful to the teams of SOHO, RHESSI, and SDO, as well as the STEREO experiment, for open access to databases. This work was supported by the Russian Foundation for Basic Research, project no. 17-02-00308. We thank Robert Sych for their assistance in the preparation of the SDO films and participation in their discussion.

21